\documentclass[final,3p,times,pdflatex]{elsarticle}
\usepackage{amssymb,bbold}
\usepackage{amsmath}
\usepackage[utf8x]{inputenc}
\usepackage{natbib}
\usepackage{graphicx}
\usepackage{hyperref}
\usepackage{epstopdf}
\usepackage{pifont}
\usepackage{float}
\usepackage{ulem}
\usepackage{soul}
\graphicspath{{pdf/}}
\hypersetup{colorlinks=true,linkcolor=blue,citecolor=blue,filecolor=blue,urlcolor=blue,breaklinks=true}
\biboptions{sort&compress}
\journal{Annals of Physics}
\RequirePackage{color}

\begin{document}

\begin{frontmatter}

\title{Modifications in the photoionization cross-section of a quantum dot with position-dependent effective mass}

\author[ufma]{Carlos Magno O. Pereira}
\ead{carlos.mop@discente.ufma.br}

\author[ufes]{Denise Assafrão}
\ead{denise.lima@ufes.br}

\author[ufma]{Frankbelson dos S. Azevedo}
\ead{frfisico@gmail.com}

\author[ceunes]{A. G. de Lima}
\ead{andre.g.lima@ufes.br}

\author[ufla]{Cleverson Filgueiras}
\ead{cleverson.filgueiras@dfi.ufla.br }

\author[ufma]{Edilberto O. Silva}
\ead{edilberto.silva@ufma.br}

\address[ufma]{Departamento de F\'{\i}sica, Universidade Federal do Maranh\~{a}o,\\ 65085-580 S\~{a}o Lu\'{\i}s, MA, Brazil}
\address[ufes]{Departmento de Física, Universidade Federal do Espírito Santo , Vitória, ES, Brazil}
\address[ceunes]{Departmento de Ciências Naturais, CEUNES, Universidade Federal do Espírito Santo , São Mateus, ES, Brazil}
\address[ufla]{Departamento de F\'{i}sica, Universidade Federal de Lavras,\\ Caixa Postal 3037, 37200-000, Lavras, Minas Gerais, Brazil}

\begin{abstract}
In this work, we investigate the photoionization cross-section of an electron confined in a quantum dot, considering the position-dependent variation of the effective mass through the parameter $\gamma$. We used a theoretical model based on the Schrödinger equation, in which $\gamma$ influences the energy levels and wave functions through an effective potential obtained from the harmonic oscillator potential—which, in the limit $\gamma = 0$, reduces to the original harmonic oscillator potential. Furthermore, we compared the modifications in the photoionization cross-section of these quantum systems with the constant-mass case. Our results demonstrate that even a small variation in $\gamma$ significantly impacts the photoionization process's amplitude and peak position. We also found that for specific values of $\gamma$, an inversion occurs: The amplitude, which initially increases as the quantum dot absorbs the photon, begins to decrease. Additionally, we observed that the optical transitions involving the ground state restrict the admissible values of $\gamma$ to negative values only. These results may have relevant implications for designing optoelectronic devices based on quantum dots with adjustable mass properties.
\end{abstract}


\end{frontmatter}

\section{Introduction}

Low-dimensional semiconductor structures exhibit unique optical properties due to their reduced dimensionality. These structures display intriguing nonlinear optical responses, such as optical limiting and saturable absorption, making them highly relevant for applications in photonics and optoelectronics \cite{https://doi.org/10.1002/adma.201605886}. The optical behavior of low-dimensional materials is influenced by various factors, including lattice dynamics, excitons, and spin-related phenomena \cite{Ogawa1996OpticalPO}. Quantum confinement effects in these systems modify electronic states and optical transitions, enabling exploration through techniques like resonant tunneling and ballistic transport measurements. Advances in fabrication techniques have led to the development of diverse low-dimensional structures, including quantum wells, wires, and dots, broadening the potential for novel optoelectronic devices \cite{chamberlain2012electronic,cryst13010108}. Among these, Quantum Dots (QDs) and Quantum Rings (QRs) have gained particular attention due to their remarkable electronic and optical properties, driven by quantum confinement effects \cite{PE.2025.165.116122,PLA.2025.130226,EPJP.2018.133.395,jacak2013quantum,PhilMag.99.2019,SR.2019.9.1427,OM.2019.91.309,Optik.2022.261.169187,EPJP.2022.137.175,SSC.2011.151.289,IEEEJQE.1987.23.2196,OC.2011.284.5818,OC.2013.291.386,rotating}.

Photoionization Cross-Sections (PCSs) play a crucial role in understanding the behavior of atoms and ions under electromagnetic radiation. In mesoscopic systems, PCSs depend on multiple factors \cite{PSSB.2002.232.209,PRB.2008.77.045317,SM.2013.58.94,SRL.2006.13.747,PLA.2023.466.128725,Atoms.2022.10.39,OLT.2025.182.111822,PB.2024.677.415717,PB.2025.696.416647}. In particular, quantum size effects and impurity position significantly influence the cross-section in GaAs QDs with infinite barriers \cite{PSSB.2002.232.209}. 
Experimental studies have employed multi-step laser excitation and ionization techniques to measure cross-sections for excited states in various atoms \cite{Atoms.2022.10.39}. Electric fields have also been shown to impact the cross-section in both finite and infinite potential barrier models \cite{IJMPB.2009.23.2127}. Studies on ZnS/SiO2 QDs reveal that parameters such as dot radius, normalized photon energy, and potential barrier height affect the cross-section \cite{PS.2012.85.045708} and oscillator strength for intersubband transitions \cite{PE.2007.36.40,PB.2015.457.126}.
Furthermore, recent work on core/shell piezoelectric QDs demonstrates that the cross-section is maximized when the impurity is centered \cite{PLA.2023.466.128725}. Some of us have published a review article and an original research paper on photoionization processes in quantum systems at the mesoscopic scale. Our findings indicate that variations in parameters such as average radius, Aharonov-Bohm flux, angular velocity, and incident photon energy significantly alter the characteristics of the process \cite{CTP.2024.76.105701}.

Position-Dependent Mass (PDM) systems have gained considerable attention in quantum mechanics due to their ability to model spatially varying material properties. In particular, Kaluza-Klein compactifications induce PDM in four-dimensional systems, leading to radially symmetric models and enabling dynamical curvature generation \cite{BALLESTEROS2017701}. PDM systems can be quantized using Killing vector fields and Noether momenta, offering a robust framework for constructing quantum Hamiltonians for nonlinear oscillators \cite{Cariñena_2017}. PDM is pivotal in semiconductor physics, especially in non-crystalline materials with large interatomic scaling \cite{ELNABULSI2020109384}. For quantum wells, PDM has a pronounced impact on nonlinear optical properties, enhancing optical absorption coefficients and refractive index changes \cite{doi:10.1142/S0217979217500096}.

Inspired by these studies, we investigate the photoionization cross-section of a QD, modeled as a system with a position-dependent effective mass. This spatial dependence accounts for material inhomogeneity in this model, which is essential for accurately describing electronic states and optical transitions. Specifically, we employ a radially dependent mass model characterized by a power-law dependence,  $m(r) = \mu r^\gamma $, where $\mu$ and $\gamma$ are nonzero real parameters. The details of this model are presented in Ref.~\cite{PhysRevA.66.042116}.

This paper is organized as follows: Section \ref{sec:PDM} provides a detailed formulation of the QD model within a position-dependent effective mass framework. We derive the eigenvalue equations, examine the confining potentials, and outline the analytical methods for determining energy levels and wavefunctions. Numerical analysis of the wavefunctions is also conducted in this section. Section \ref{PCS} focuses on the photoionization process, offering an in-depth discussion on how position-dependent mass influences this optical phenomenon. Numerical results are presented and analyzed, highlighting the effects of various parameter values on the optical responses of QDs. Finally, Section \ref{conc} summarizes the key findings and discusses the potential applications of these results in designing optoelectronic devices based on quantum systems with PDM. 

\section{\label{sec:PDM}Position-Dependent Effective Mass Model}

In this section, we present a model describing the motion of an electron within a QD, where the effective mass varies with position. This approach allows for a detailed exploration of the effects of spatial mass variation on the electronic and optical properties of the system. The eigenvalue equation is derived by introducing a radial confining potential and ensuring the Hermiticity of the Hamiltonian, as required by the noncommutative relationship between the momentum and PDM operators \cite{LIMA2023115688,dekar}. 

We start with the Schrödinger equation for a particle whose mass depends on the position \cite{PhysRevA.66.042116,Alhaidari2003}
\begin{equation}
-\frac{\hbar ^{2}}{2}\boldsymbol{\nabla} \frac{1}{m(r)}\boldsymbol{\nabla} \Psi (r,\theta ,\phi
)+V(r)\Psi (r,\theta ,\phi )=E\Psi (r,\theta ,\phi ).  \label{scr}
\end{equation}
By proposing the separable solution
\begin{equation}
\Psi (r,\theta ,\phi )=\frac{f(r)}{r}Y_{\ell }^{m}(\theta ,\phi ),
\label{sb}
\end{equation}
we get two equations
\begin{equation}
f''(r) -\frac{\ell (\ell +1)}{r^{2}}f\left(
r\right) +\frac{m'(r)}{m(r)}\left( \frac{1
}{r}-\frac{d}{dr}\right) f(r) -\frac{2m(r)}{\hbar ^{2}}\left(
V(r) -E\right) f(r) =0,\label{radmpd}
\end{equation}
where $m'(r) = dm(r)/dr$, and
\begin{equation}
    \mathbf{L}^2Y_{\ell}^{m}(\theta,\phi)= \hbar^2 \ell (\ell +1)Y_{\ell}^{m}(\theta, \phi),\label{eqangular}
\end{equation}
with $\ell=0,1,2,\dots$ being the angular momentum quantum number for $-\ell \leq m \leq +\ell$ and $m$ characterize the z component of the angular momentum $L_{z}\geq m \hbar$. The solutions to the angular differential equation (\ref{eqangular}), where the operator  $\mathbf{L}^2$  is defined as 
\begin{equation}
\mathbf{L}^2 = -\hbar^2 \left[ \frac{1}{\sin\theta} \frac{\partial}{\partial \theta} \left( \sin\theta \frac{\partial}{\partial \theta} \right) + \frac{1}{\sin^2\theta} \frac{\partial^2}{\partial \phi^2} \right],
\end{equation}
are the well-known spherical harmonics  $Y_{\ell}^{m}(\theta, \phi)$. 

Equation  (\ref{radmpd}) accounts for the coupling between the PDM and the potential energy. Also, the radial function  $f_{\ell}(r)$ depends on the total angular-momentum quantum number $\ell$. We will soon see that the requirement of square-integrability for the wave function introduces an additional quantum number, the radial quantum number $n$.

The scalar potential $ V(r) $ and the PDM function $ m(r) $ are respectively given by
\begin{equation}
V(r) = \frac{1}{2}m(r)\, \omega_0^2\, r^2, \label{radp}
\end{equation}
\begin{equation}
m(r) = \mu r^\gamma, \label{massf}
\end{equation}
where $ \omega_{0} $ characterizes the transverse confinement frequency, and $ \mu $ denotes the effective mass of the electron. The parameter $\gamma $ in Eq. (\ref{massf}) controls the magnitude and spatial variation of the mass. This model captures the inherent inhomogeneity in many semiconductor quantum structures, such as quantum wells, wires, and dots \cite{Bastard1988,PLA.2018.382.2868}. This inhomogeneity arises due to the spatial variation in material composition or confinement conditions, which is common in semiconductor heterostructures \cite{altshuler2012mesoscopic}. The parameter $\gamma$ governs the magnitude and spatial dependence of the mass. For example, $\gamma = 0$ corresponds to a constant mass, while $\gamma \neq 0$ describes a mass that varies with the radial distance $r$. Additionally, it accounts for interface effects in heterostructures, where the effective mass changes at material boundaries \cite{PLA.2000.275.25}. The interplay between the PDM and the potential $V(r)$ leads to modifications in the system's energy spectrum and transport properties, making it a versatile tool for studying inhomogeneous quantum systems \cite{PRA.1999.60.4318}. Here, we focus on a GaAs sample with an effective electron mass of $\mu = 0.067 \,m_{e}$, where $m_{e}$ is the electron mass in vacuum.
\begin{figure}[!h]
    \centering
    \includegraphics[scale=0.35]{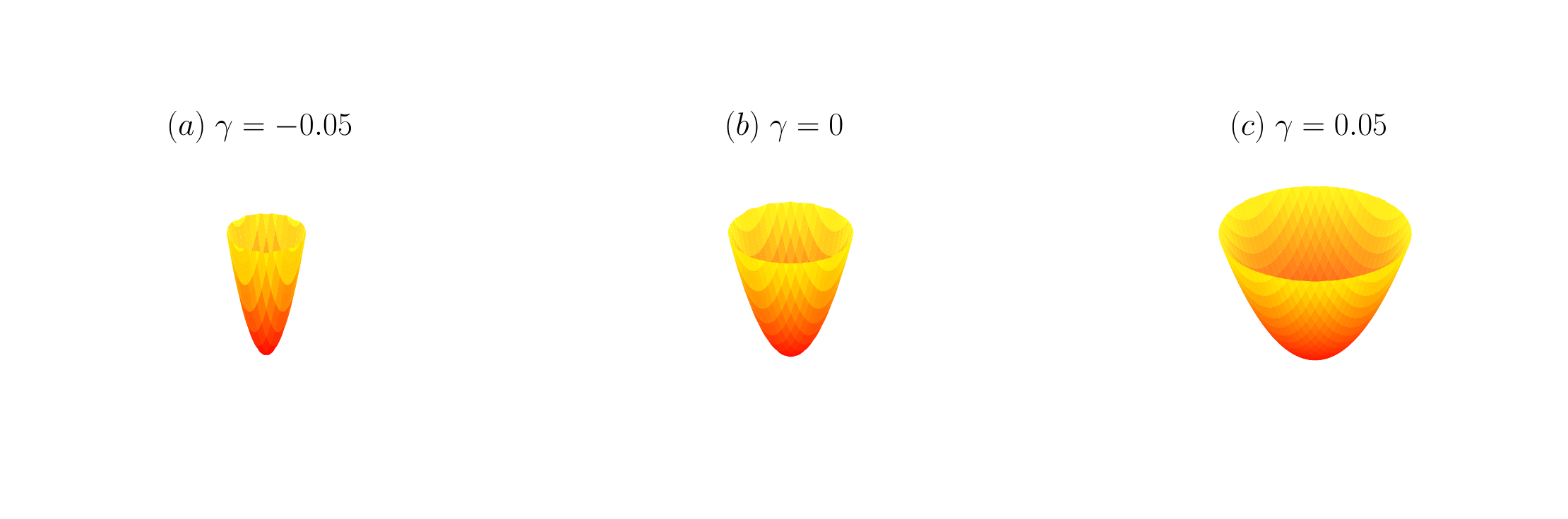}
    \caption{(Color online) Plots of the oscillator potential defined by Eq. \eqref{radp} with a position-dependent mass, considering $\hbar\omega_0 = 30\,\mathrm{meV}$, for different values of $\gamma$: (a) $\gamma = -0.05$, (b) $\gamma = 0$, and (c) $\gamma = 0.05$.}
    \label{fig:V3}
\end{figure}
A quadratic function proportional to $r^2$ describes the harmonic oscillator potential in the standard case. The inclusion of a term associated with the mass modifies the potential to a dependence of the form $r^{\gamma + 2}$, altering its nature for  $\gamma \neq 0$  and, consequently, the type of confinement to which the particle will be subjected. Figure \ref{fig:V3} illustrates the influence of the parameter $\gamma$ on the oscillator potential. We observe that as $\gamma$ increases, the concavity of the potential becomes wider, indicating a reduction in the confinement strength. The parameter $\gamma$ causes modifications in the confinement potential, and its value can either increase or decrease the confining force. Figure \ref{fig:V3} shows three graphs of the potential, according to Eq. \eqref{radp}, for different cases: with $\gamma$ negative, zero, and positive. In Fig. \ref{fig:V3}(a), with $\gamma = -0.05$, the mass decreases as the distance increases, resulting in a more pronounced confinement. Fig. \ref{fig:V3}(b) illustrates the standard case, where the mass remains constant ($\gamma = 0$). Finally, in Fig. \ref{fig:V3}(c), with $\gamma = 0.05$, the mass increases with distance, making the concavity of the potential broader and reducing the confinement.

\begin{figure}[!h]
    \centering
    \includegraphics[scale=0.35]{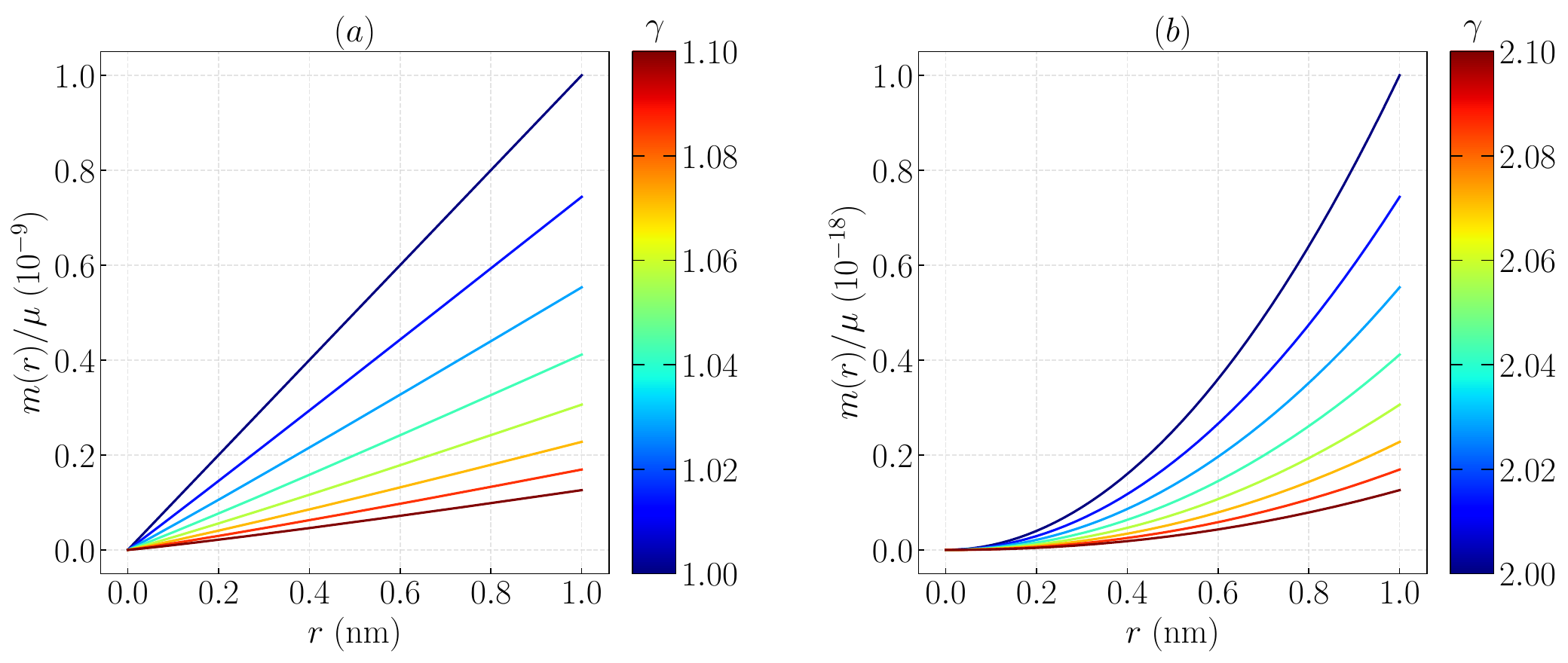}
    \caption{(Color online) Plots of the ratio between mass and effective mass of the electron as a function of position for different values of $\gamma$: in (a), $\gamma$ varies from 1.0 to 1.1; in (b), from 2.0 to 2.1.}
    \label{fig:mass}
\end{figure}

The position-dependent effective mass undergoes significant variations even for small changes in the parameter $\gamma$. Fig. \ref{fig:mass} shows the behavior of the ratio between the mass and effective mass of the electron. In (a), as the parameter $\gamma$ increases, the behavior is linear but with a decreasing rate of change. However, in (b), we observe a quadratic behavior with decreasing concavity as $\gamma$ grows. The effect of these features will be analyzed in the following sections of this work.

Since we are interested in studying the model's effective potential, we must eliminate the first derivative term  $f'(r)$  in Eq. (\ref{radmpd}). For this to be accomplished, we use the transformation
\begin{equation}
f(r) = g(r) e^{\int h(r) \, dr}, \label{tr}
\end{equation}
where $h(r)$  is a function to be found to cancel the first derivative term. We obtain the equation 
\begin{equation}
g''(r)+ g'(r) \left[ 2 h(r) + \frac{m'(r)}{m(r)} \right] + g(r) \left[ h^2(r) + h'(r) - \frac{\ell(\ell + 1)}{r^2} - \frac{m'(r)}{m(r)} \frac{1}{r} + \frac{2m(r)}{\hbar^2} \left(V(r) - E\right) \right] = 0. \label{neq}
\end{equation}
By setting $2 h(r) + m'(r)/m(r) = 0$, we find 
$h(r) = -m'(r)/2m(r)$. In possession of this result and using the equations (\ref{radp}) and (\ref{massf}), Eq. (\ref{neq}) becomes 
\begin{equation}
-\frac{\hbar^2}{2 \mu}g''(r) + \left(V_{\text{eff}}(r) - E\right) g(r) = 0, \label{radial}
\end{equation}
where
\begin{equation}
V_{\text{eff}}(r) =\frac{\hbar^2}{2 \mu} \frac{\gamma \left( \gamma -2\right) +4\ell \left( \ell
+1\right)}{4r^{2}}+\frac{\mu \omega_{0}^{2}}{2}r^{2\gamma
+2}-E r^{\gamma}
\end{equation}
\begin{figure}[!h]
\centering
\includegraphics[scale=0.35]{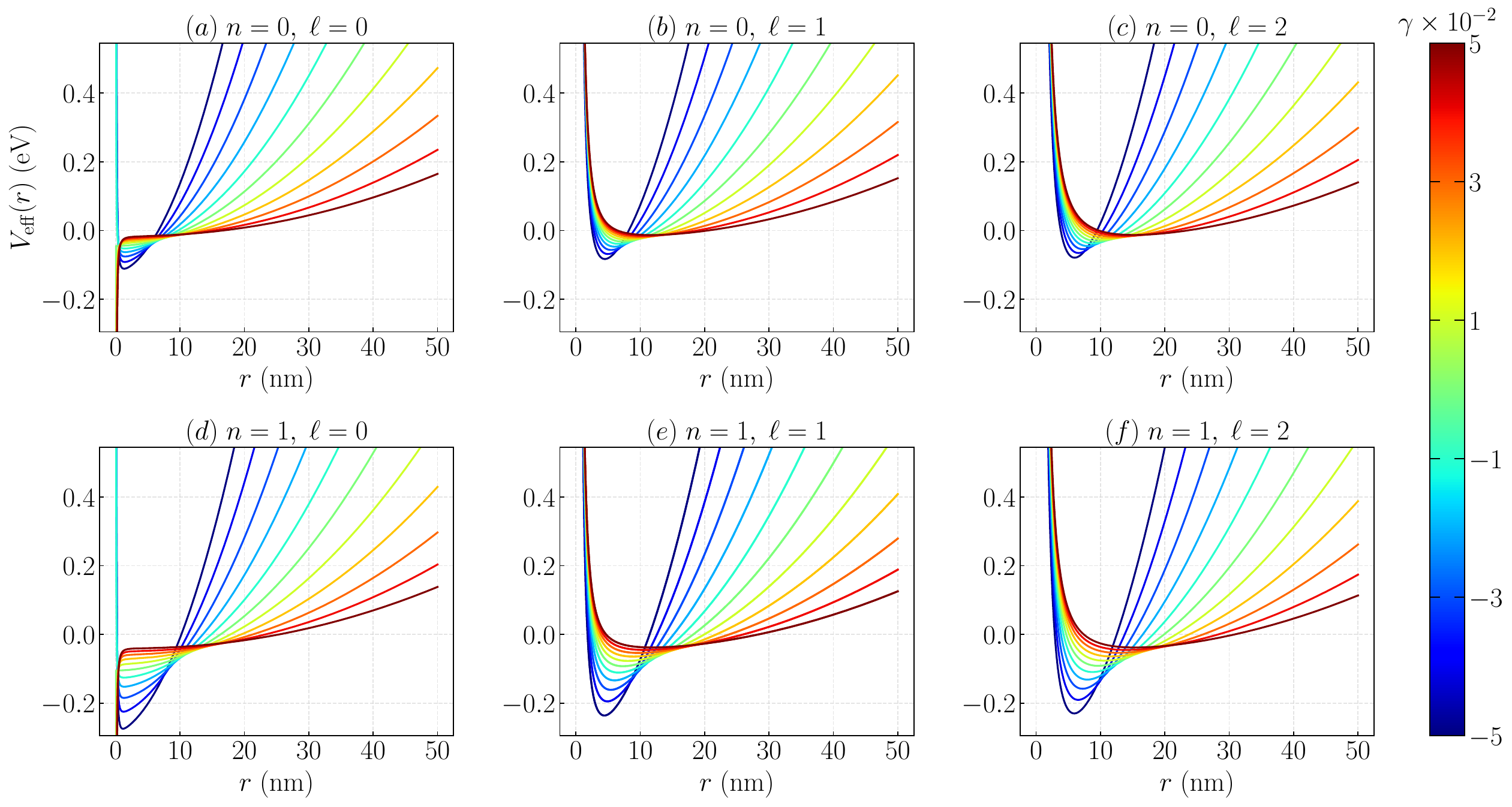}
\caption{(Color online) The Figures (a), (b), (c), (d), (e), (f) depict the effective potential for fixed values of $\hbar\omega_{0} = 30$ meV, corresponding to quantum states $(n=0, \ell=0)$, $(n=0, \ell=1)$ , $(n=0, \ell=2)$ , $(n=1, \ell=0)$, $(n=1, \ell=1)$ and $(n=1, \ell=2)$, respectively, with the parameter $\gamma$ varying from -0.05 to 0.05 in intervals of 0.01.}
\label{veff}
\end{figure}
is the effective potential. The effective potential $V_{\text{eff}}(r)$ combines three main contributions: a centrifugal term proportional to $1/r^2$, a radial term proportional to $r^{2\gamma + 2}$, and a linear term proportional to $r^\gamma$, each of which significantly affects the system's behavior in different radial regimes (Fig. \ref{veff}). Near the origin ($r \to 0$), the centrifugal term dominates, leading to $V_{\text{eff}}(r) \to \infty$ for $\ell > 0$, which corresponds to a centrifugal barrier (Fig. \ref{veff}(b)). For $\ell = 0$, the potential is less singular. In intermediate radial regions, the interplay between the centrifugal term, the radial confinement term, and the linear energy-dependent term defines the depth and position of potential wells, where the wave functions are likely to be localized. At large distances ($r \to \infty$), the radial term $r^{2\gamma + 2}$ dominates, resulting in an asymptotic growth of the potential, ensuring the confinement of the wave function.

The parameter $\gamma$ significantly impacts the shape of the effective potential, $V_{\text{eff}}(r)$. For $\gamma = 0$, the radial term reduces to a parabolic form ($\propto r^2$), while increasing $\gamma$ enhances the confinement strength at large distances and modifies the potential profile in intermediate regions due to the contributions from the $r^{\gamma}$ term. Conversely, when $\gamma < 0$, the radial term $r^{2\gamma + 2}$ decreases more slowly as $r \to \infty$, leading to weaker confinement at large distances (Fig. \ref{veff}). Depending on the interplay with other terms, this can result in more delocalized wave functions. The angular momentum quantum number $\ell$ introduces a centrifugal barrier that increases with higher $\ell$, shifting the potential minimum to larger values of $r$. The eigenenergy $E$ further influences the potential; positive energies lower the effective potential in the intermediate region through the $-Er^\gamma$ term, whereas negative energies increase the effective potential, leading to stronger confinement.

The effective potential exhibits distinct behavior depending on the quantum numbers considered. Figure \ref{veff} illustrates two sets of cases: in the first, with \( n = 0 \), the potentials corresponding to \( (n=0,\,\ell=0) \), \( (n=0,\,\ell=1) \), and \( (n=0,\,\ell=2) \) are presented in figures (a), (b), and (c), respectively; in the second, with \( n = 1 \), the cases \( (n=1,\,\ell=0) \), \( (n=1,\,\ell=1) \), and \( (n=1,\,\ell=2) \) are shown in (d), (e), and (f). 
It is noted that, in the second set (\( n = 1 \)), the potential minima are more widely spaced for each value of \( \gamma \), indicating that the effect of this parameter is more pronounced compared to the first set (\( n = 0 \)). Higher values of $\ell$ result in a more pronounced centrifugal barrier, shifting the potential minimum to farther positions. In other words, for larger values of $\gamma$, the potential becomes weaker at large distances, reducing confinement. On the other hand, for smaller values of $\gamma$, the weakened confinement may allow states with broader spatial distributions, especially for lower angular momentum. The energy $E$ controls the overall potential profile, with positive values reducing the intermediate potential and negative values increasing confinement. These qualitative features are consistent with the observed numerical results, as shown in the provided figures. The shape and depth of the potential wells and the asymptotic behavior are in excellent agreement with the model's predictions. This analysis highlights the versatility of $V_{\text{eff}}(r)$ in describing spatially dependent mass systems and their capacity to capture the intricate interplay of confinement and angular momentum effects.
It is observed in Figures (a) and (d) that, for \( \ell = 0 \), bound states exist only for \( \gamma \leq 0 \) and do not form for positive values of \( \gamma \). This is an important point when analyzing the PCS, which follows a selection rule for the values of $\ell$. This restriction implies that only negative values of $\gamma$ can be considered when $\ell =0$

\begin{figure}[!h]
    \centering
    \includegraphics[scale=0.35]{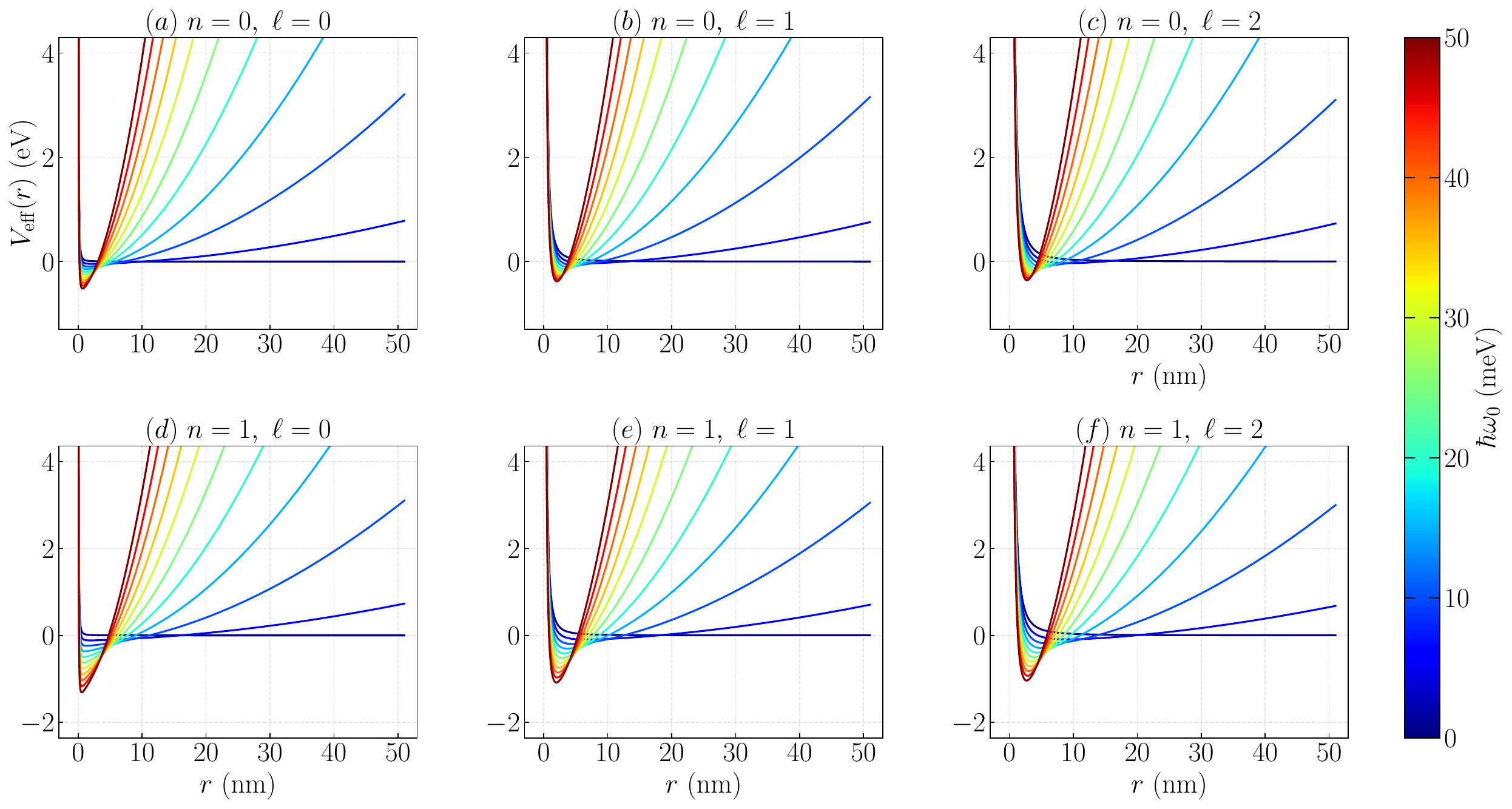}
    \caption{(Color online) The figures (a), (b), and (c) show the effective potential for the quantum states $(n=0, \ell=0)$, $(n=0, \ell=1)$, and $(n=0, \ell=2)$. Meanwhile, figures (d), (e), and (f) display the effective potential for the states $(n=1, \ell=0)$, $(n=1, \ell=1)$, and $(n=1, \ell=2)$. The graphs represent the effective potential as a function of $r$, considering $\hbar \omega_0$ varying from 0 to $50 ~ \mathrm{meV}$ in increments of $5~\mathrm{meV}$, with $\gamma=-0.1$.}
    \label{veff-omega}
\end{figure}
The confinement energy $\hbar\omega_{0}$ plays a crucial role in the effective confinement potential, as it determines the magnitude of the confinement force. Fig. \ref{veff-omega} presents the effective potential profiles as a function of $ r $ for $ \hbar\omega_{0} $ values ranging from $ 0 $ to $ 50~\mathrm{meV} $. In the trivial case of $ \hbar\omega_{0} = 0 $, there is no confinement. As $ \hbar\omega_{0} $ increases, the confinement becomes stronger. Fig. \ref{veff-omega} shows the plots of the effective potential as a function of position for different values of the quantum numbers \((n, \ell)\). In plots (a), (b), and (c), corresponding to \((n=0, \ell=0)\), \((n=0, \ell=1)\), and \((n=0, \ell=2)\), the potential minima shift to larger \(r\) values as \(\ell\) increases. In plots (d), (e), and (f), which represent \((n=1, \ell=0)\), \((n=1, \ell=1)\), and \((n=1, \ell=2)\), the minima become less energetic. All plots were generated considering \(\gamma = -0.1\), a negative value chosen to create a potential well suitable for the quantum number configurations analyzed. 

\begin{figure}[!h]
    \centering
    \includegraphics[scale=0.35]{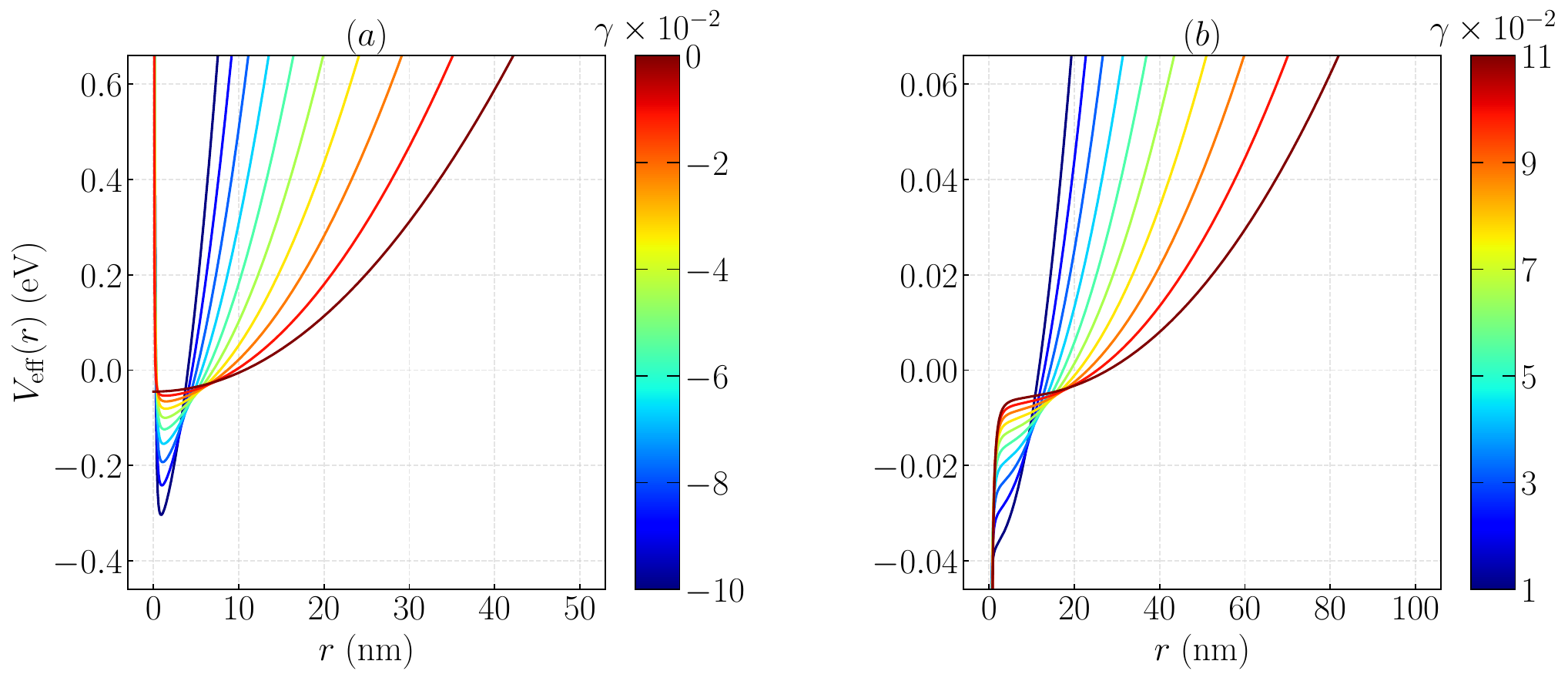}
    \caption{(Color online) Figure (a) shows the effective potential as a function of $ r $ for negative $ \gamma $ values ranging from $-0.11$ to $-0.01$, while Figure (b) presents the effective potential for positive $ \gamma $ values in the range of $ 0 $ to $ 0.10 $. In both cases, $ \hbar\omega_{0} = 30~\mathrm{meV} $ was considered. The analyzed configuration corresponds to the state $ (n=0, \ell=0) $, resulting in a bound state only in case (a).}
    \label{fig:veff00}
\end{figure}
 Not all combinations of the quantum numbers $(n, \ell)$ and $\gamma$ result in bound states. When $\ell = 0$, there is a restriction on the values of $\gamma$. Fig. \ref{fig:veff00} shows two graphs of the effective potential as a function of position for different values of $\gamma$, with $\gamma$ taking only negative values in graphs (a) and (b). In plot (a), with $(n=0, \ell=1)$, a potential well is observed, while in plot (b), with $(n=0, \ell=0)$, no well is formed. These results demonstrate that for $\ell = 0$ and $\gamma > 0$, no potential well is formed, and thus, there is no confinement. For $\gamma = 0$, the potential reduces to that of the standard harmonic oscillator. For this reason, we consider the transition $\psi_{11} \rightarrow \psi_{00}$ only for negative values of $\gamma$. We also include the transition $\psi_{02} \rightarrow \psi_{01}$, since in this configuration, bound states exist for any value of $\gamma$, except when $\gamma = -2$.

The eigenvalues and eigenfunctions of the radial equation (\ref{radial}) are
\begin{equation}
g_{n,\ell}(r)=a_{n}\,r^{-\frac{\gamma +1}{2}}\xi^{\nu}M\left( -n,\lambda ,\xi \right) e^{-\xi},
\end{equation} 
where $M(-n,\lambda,\xi)$ is the confluent hypergeometric function of the first kind, $a_{n}$ is the normalization constant, and
\begin{equation}\label{eq:Enm}
E=\frac{\gamma +2}{2}\left[ 2n+1+\frac{\sqrt{(\gamma -1)^{2}+4\ell(\ell+1)}}{
\gamma +2}\right]\hbar \omega_{0},
\end{equation}
where we have defined the following parameters: 
\begin{align}
\xi & =\frac{2 \mu \omega _{0}r^{\gamma +2}}{\hbar (\gamma +2)}, \;\;
\nu = \frac{\sqrt{(\gamma - 1)^2 + 4\ell(\ell+1)}}{2(\gamma + 2)} + \frac{1}{2}, \;\;
\lambda = 1 + \frac{\sqrt{(\gamma - 1)^2 + 4\ell(\ell+1)}}{\gamma + 2}.
\end{align}

The function (\ref{tr}) can be written as 
\begin{equation}
f_{n,\ell}(r) = g_{n,\ell}(r) \,e^{\int \frac{m'(r)}{2m(r)} dr},
\label{trs}
\end{equation}
which using (\ref{massf}) yields
\begin{equation}
f_{n,\ell}(r)=g_{n,\ell}(r)\,r^{\frac{\gamma}{2}}. \label{tr2}
\end{equation}
Therefore, the solution (\ref{sb}) is written as
\begin{equation}
\Psi_{n,\ell,m} (r,\theta ,\phi )=C_{n,\ell} \,g_{n,\ell}(r)\,r^{\frac{\gamma}{2}-1}Y_{\ell}^{m}(\theta,\phi ),
\end{equation}
where $C_{n,\ell}$ is the normalization constant, 
\begin{equation}
g_{n,\ell}(r) =
r^{-\frac{\gamma +1}{2}}
\left( \frac{2 \mu \omega_0}{\hbar (\gamma + 2)}r^{\gamma+2} \right)^{\frac{1}{2(\gamma + 2)} \sqrt{(\gamma - 1)^2 + 4l(l+1)} + \frac{1}{2}}
e^{-\frac{\mu \omega_0}{\hbar (\gamma + 2)}r^{\gamma+2}}\,M\left(-n, \frac{\gamma + 2 + \sqrt{(\gamma - 1)^2 + 4l(l+1)}}{\gamma + 2}, \frac{2 \mu \omega_0}{\hbar (\gamma + 2)}r^{\gamma+2}\right)    
\end{equation}
and 
\begin{equation}
Y_l^m(\theta, \phi) = (-1)^m \sqrt{\frac{(2l + 1)}{4\pi} \frac{(l - m)!}{(l + m)!}} P_l^m(\cos\theta) e^{im\phi},   
\end{equation}
with $P_\ell^m(x)$ being the associated Legendre polynomials.

\begin{figure}[!h]
\centering
\includegraphics[scale=0.35]{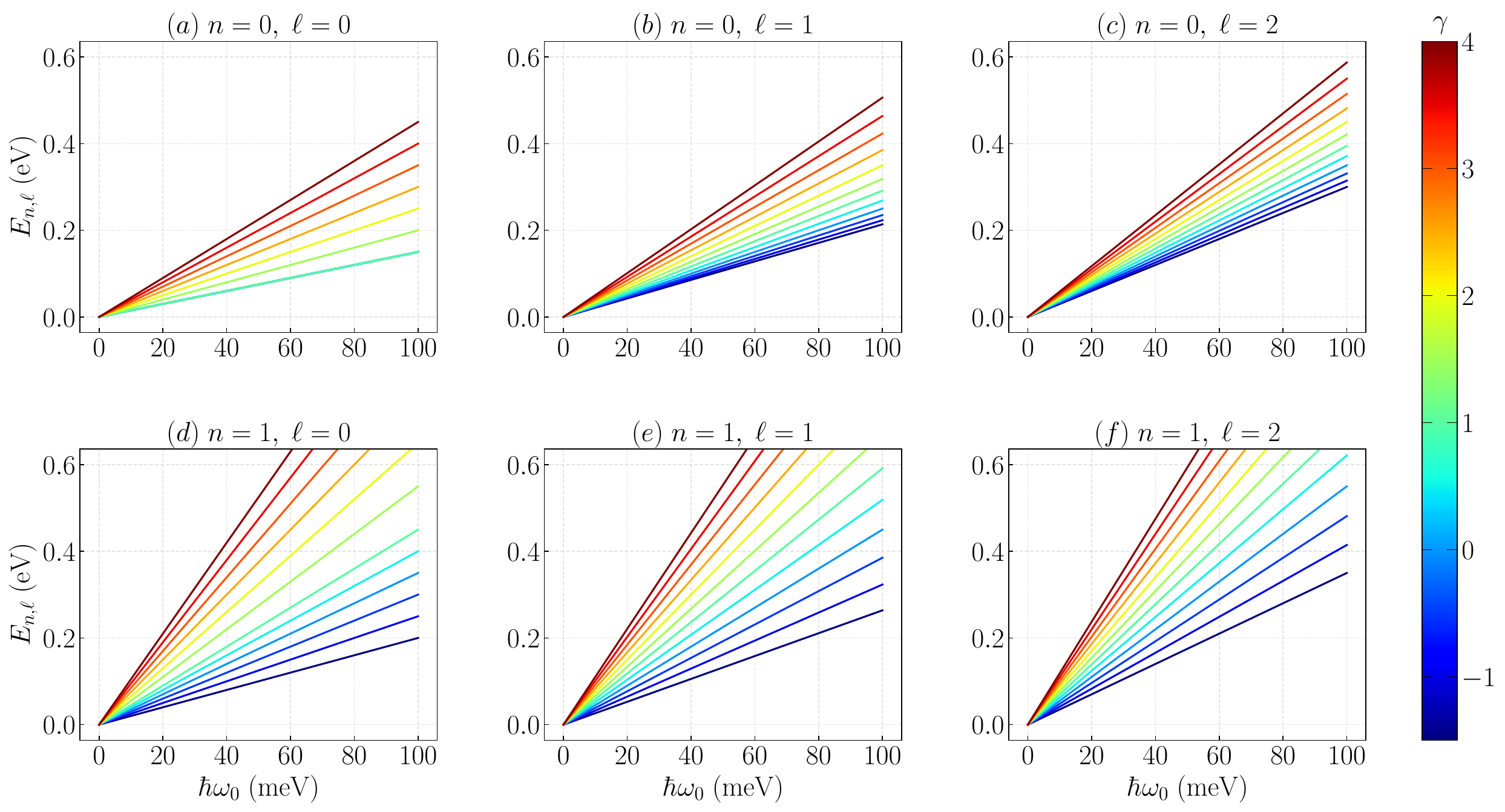}
\caption{(Color online) The figures show the energy levels as a function of \(\hbar \omega_0\) for positive and negative values of the parameter \(\gamma\). Figures (a), (b), and (c) display the quantum states \((n = 0,\,\ell=0)\), \((n = 0,\,\ell=1)\), and \((n = 0,\,\ell=2)\). Meanwhile, Figures (d), (e), and (f) present the states \((n = 1,\,\ell=0)\), \((n = 1,\,\ell=1)\), and \((n = 1,\,\ell=2)\).}
\label{fig:energy}
\end{figure}

The equation \eqref{eq:Enm} describes the variation of energy eigenvalues as a function of the parameter $\gamma$, highlighting that it cannot take the value $-2$. Figure \ref{fig:energy} presents the energy profiles as a function of $\hbar \omega_0$ for different values of $\gamma$. It is observed that the energy curves exhibit a linear behavior, with slopes determined by $\gamma$: larger values of $\gamma$ result in steeper slopes.

This linear behavior occurs for all analyzed quantum states. Figures \ref{fig:energy}(a), \ref{fig:energy}(b), and \ref{fig:energy}(c) illustrate the cases $(n=0, \ell=0)$, $(n=0, \ell=1)$, and $(n=0, \ell=2)$, respectively. For $n=1$, Figures \ref{fig:energy}(d), \ref{fig:energy}(e), and \ref{fig:energy}(f) correspond to the states $(\ell=0)$, $(\ell=1)$, and $(\ell=2)$.

In the cases with $n=1$, the separation between the curves becomes more pronounced as $\hbar\omega_{0}$ increases, compared to $n=0$. Although this separation also increases for $n=0$, the growth is significantly more pronounced for $n=1$.

\begin{figure}[h!]
    \centering
    \includegraphics[scale=0.35]{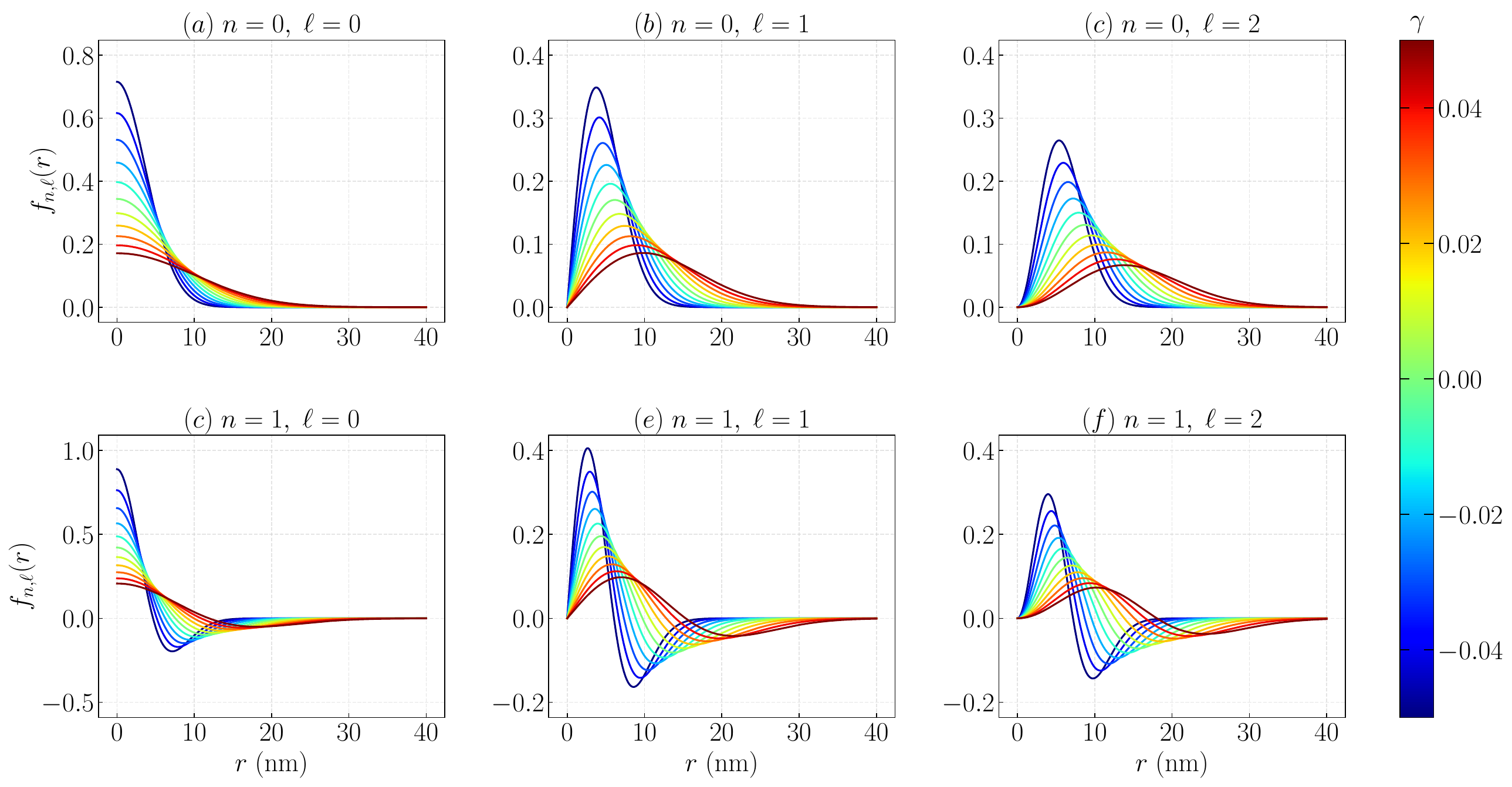}
    \caption{(Color online) Graphs of the normalized wave function $f_{n, \ell}(r)$ as a function of the radial distance $r$ (in nm) for the quantum numbers $n=0, \ell=1, n=0, \ell=2, n=1, \ell=1$, and $n=1, \ell=2$, considering values of $\gamma$ ranging from -0.05 to 0.05 . The physical parameters used in the calculations are $\mu=0.067\, m_e$ and $\hbar \omega_0=30\, \mathrm{meV}$.}
    \label{fig:wave-gamma-positive}
\end{figure}

The position-dependent mass also modifies the wave functions and probability densities, consequently affecting the shape of the transition matrix between quantum states. Fig. \ref{fig:wave-gamma-positive} shows the graph of the normalized radial wave function $\Psi(r)$ for values of $\gamma$ ranging from $-0.05$ to $0.05$ in various quantum states: ($n=0, \ell=0$), ($n=0, \ell=1$), ($n=0, \ell=2$), ($n=1, \ell=0$), ($n=1, \ell=1$) and ($n=1, \ell=2$). Near the origin, the wave amplitude is larger for smaller values of $\gamma$. However, along the radial coordinate, there is a point where this behavior reverses. As the radial quantum numbers increase, these reversals occur more frequently due to the growing number of nodes.

\begin{figure}[!h]
\centering
\includegraphics[scale=0.35]{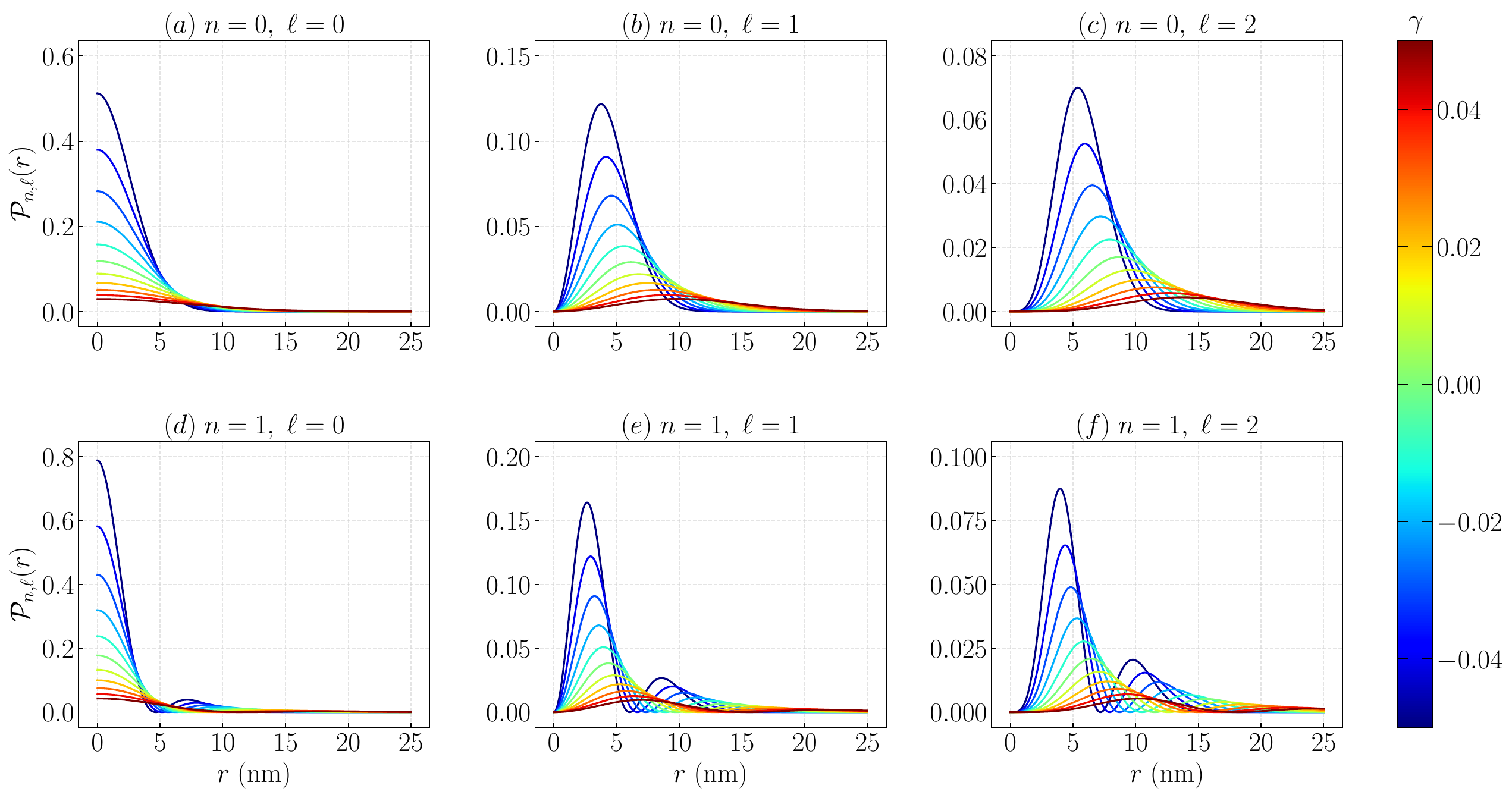}
\caption{
(Color online) 
Plots of the normalized wave function $f_{n,\ell}(r)$ (Fig. (a)) and the normalized probability density $\mathcal{P}_{n,\ell}(r)$ (Fig. (b)) as functions of the radial distance $r$ (in nm) for three different values of the parameter $\gamma = 0.0$, $\gamma = 0.01$, and $\gamma = 0.02$, with the quantum number fixed at $n = 2$. 
The wave functions and probability densities are normalized over the radial domain. 
The physical parameters used are $\mu = 0.067\, m_e$  and $\hbar \omega_0 = 30 \,\text{meV}$. 
Different colors represent the distinct values of $\gamma$.}
\label{fig:prob}
\end{figure}

Fig. \ref{fig:prob} displays the probability density $\mathcal{P}_{n,\ell}(r)=|f_{n,\,\ell}(r)|^{2}$ for the quantum states previously mentioned, illustrating the probability of finding the particle in different positions along the radial coordinate. The plot indicates that the particle is most likely to be found near the center of the quantum system, with higher probabilities associated with smaller values of $\gamma$.

\begin{figure}[!t]
\centering
\includegraphics[scale=0.4]{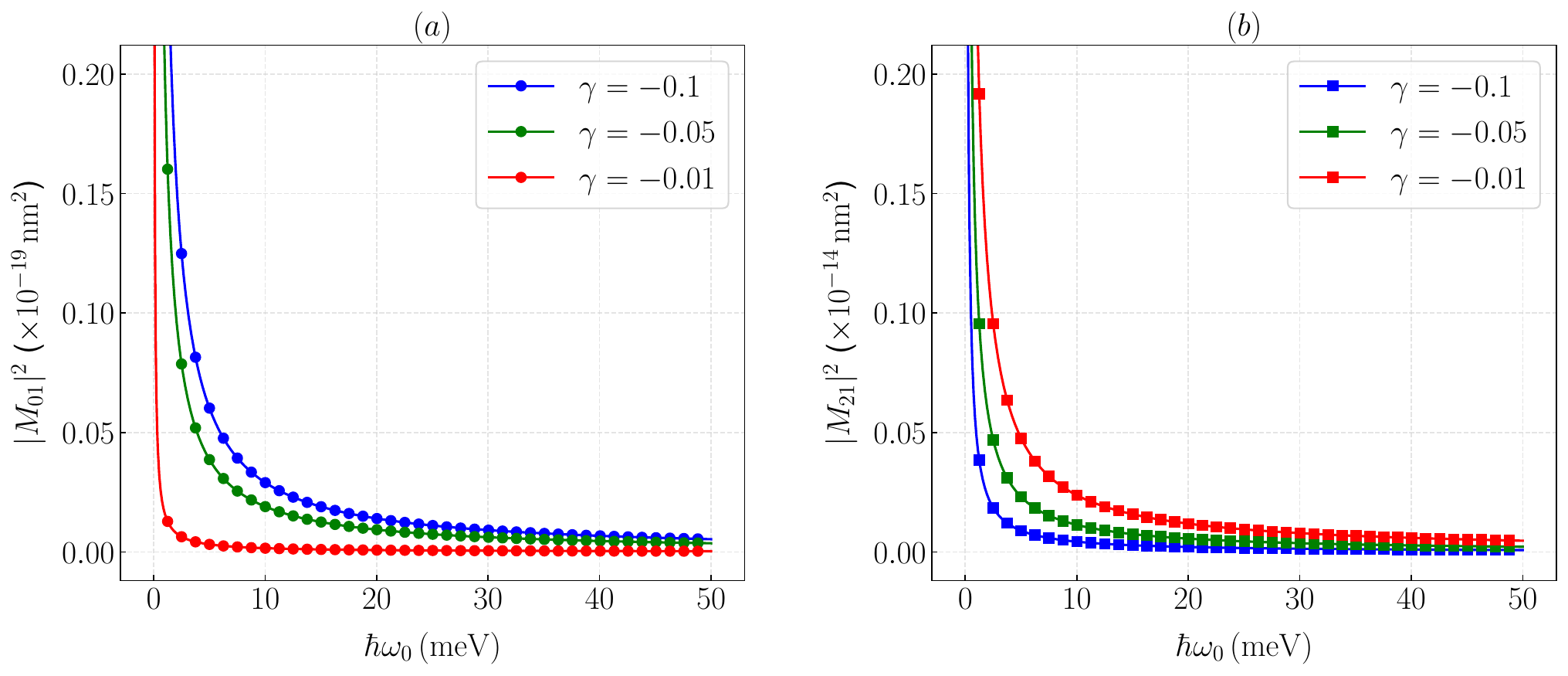}
\caption{Plot of the transition matrix element for two distinct transitions, considering negative values of $\gamma$. In (a), the transition occurs between the states $(n=0, \ell=0)$ and $(n=1, \ell=1)$. In (b), the transition involves the states $(n=0, \ell=1)$ and $(n=0, \ell=2)$.}
\label{fig:matrixM}
\end{figure}
\begin{figure}[!t]	
\centering
\includegraphics[scale=0.35]{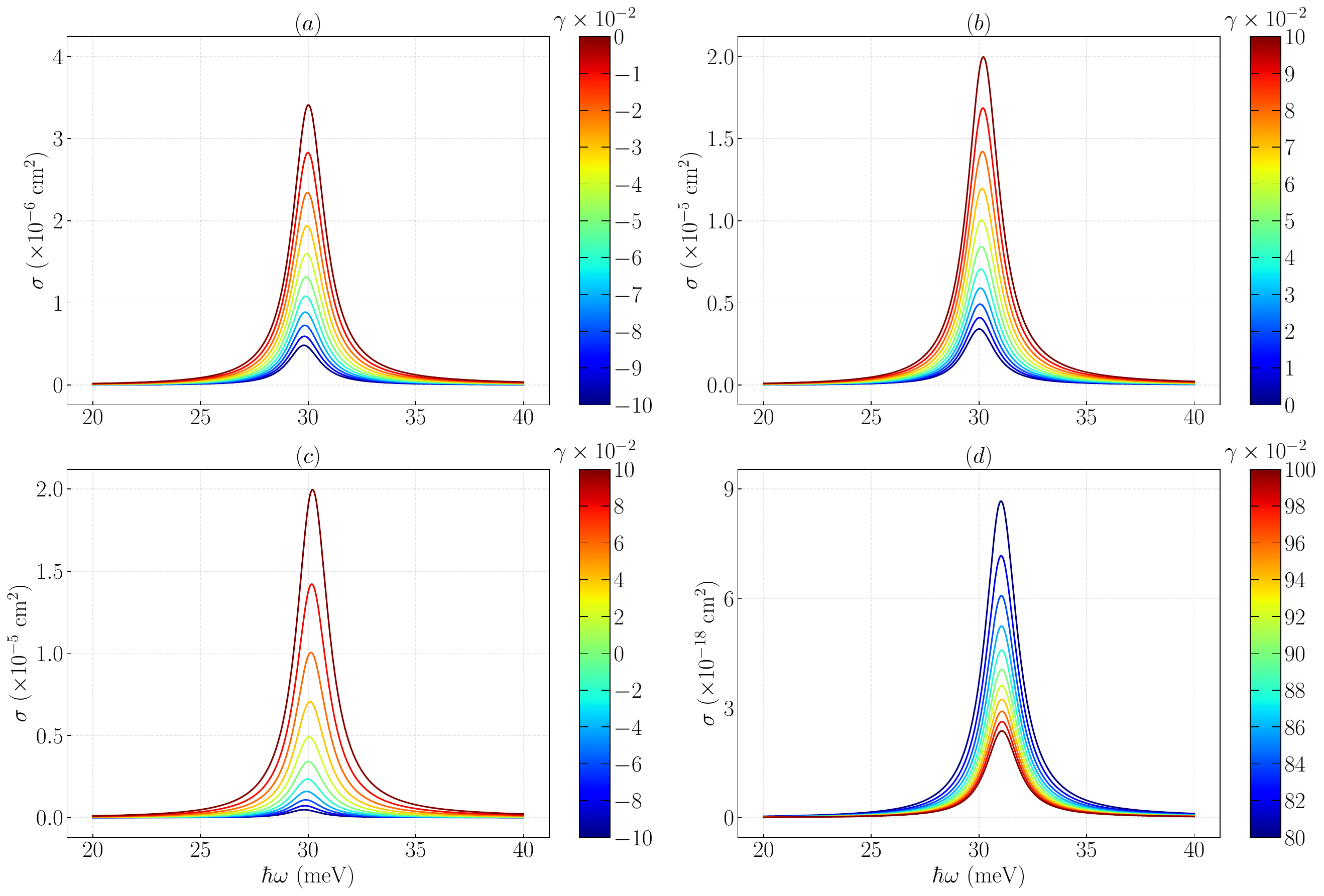}
\caption{Photoionization cross section for values of $\gamma$ on transition $(n=0,\,\ell=1)$ to $(n=0,\,\ell=2)$. In (a), $\gamma$ varies from -10 to 0; in (b), from 0 to 10; in (c), from -10 to 10, incorporating the ranges of (a) and (b); and in (d), from 80 to 100. In all cases, the values of $\gamma$ are on the order of $10^{-2}$. The physical parameters used are $\mu = 0.067 \,m_e$ and $\hbar \omega_0 = 30\,\text{meV}$.}
\label{fig:sigma1}
\end{figure}
\begin{figure}[!t]
\centering
\includegraphics[scale=0.44]{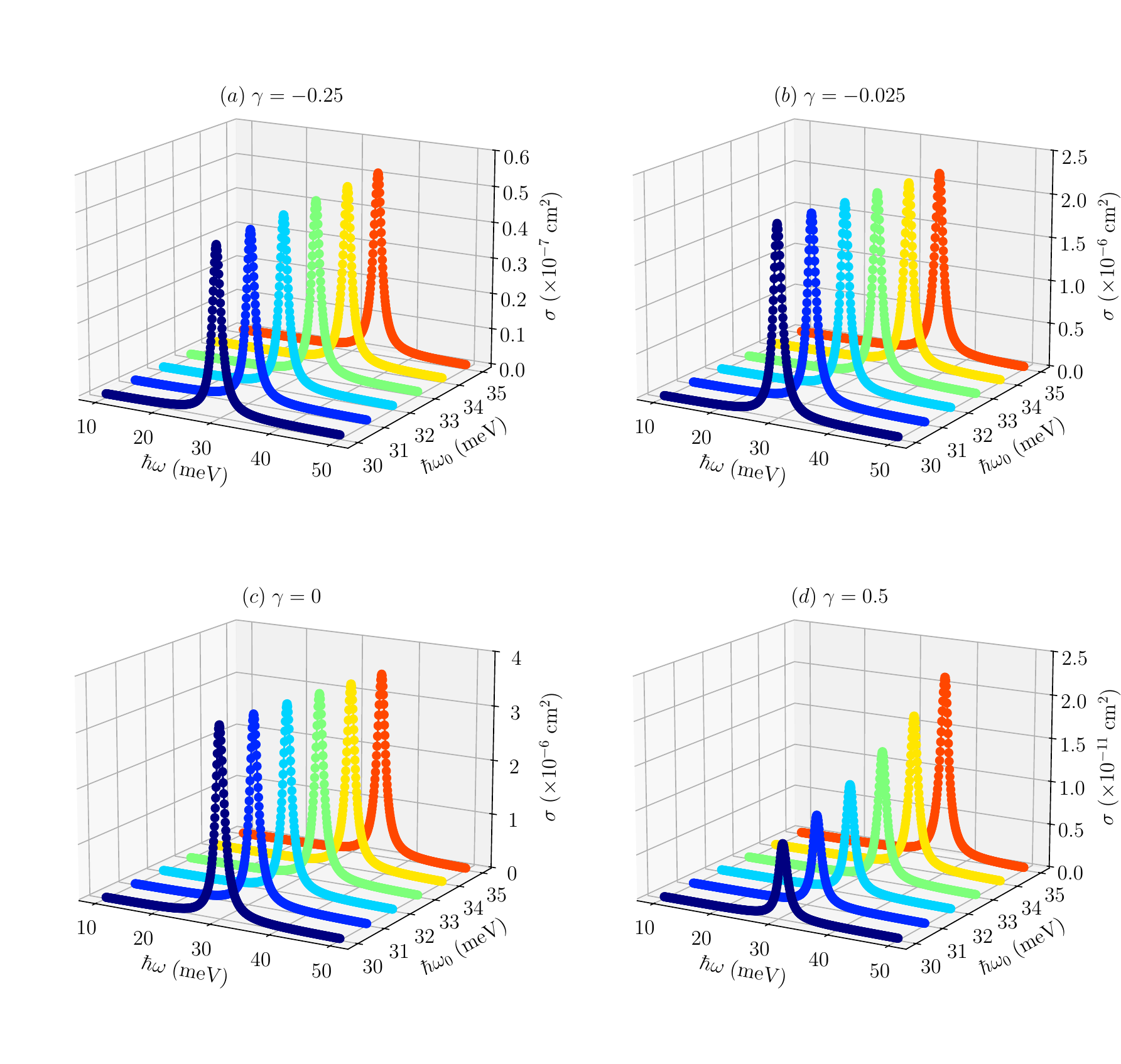}
\caption{Photoionization cross section for different values of $\hbar \omega_0$ on transition $(n=0,\,\ell=1)$ to $(n=0,\,\ell=2)$. The physical parameters used are $\mu=$ $0.067\, m_e$ and (a) $\gamma=-0.25$, (b) $\gamma=-0.025$, (c) $\gamma=0$ and (d) $\gamma=0.5$.}
\label{fig:pcs3d0102}
\end{figure}
\begin{figure}[!t]	
\centering
\includegraphics[scale=0.35]{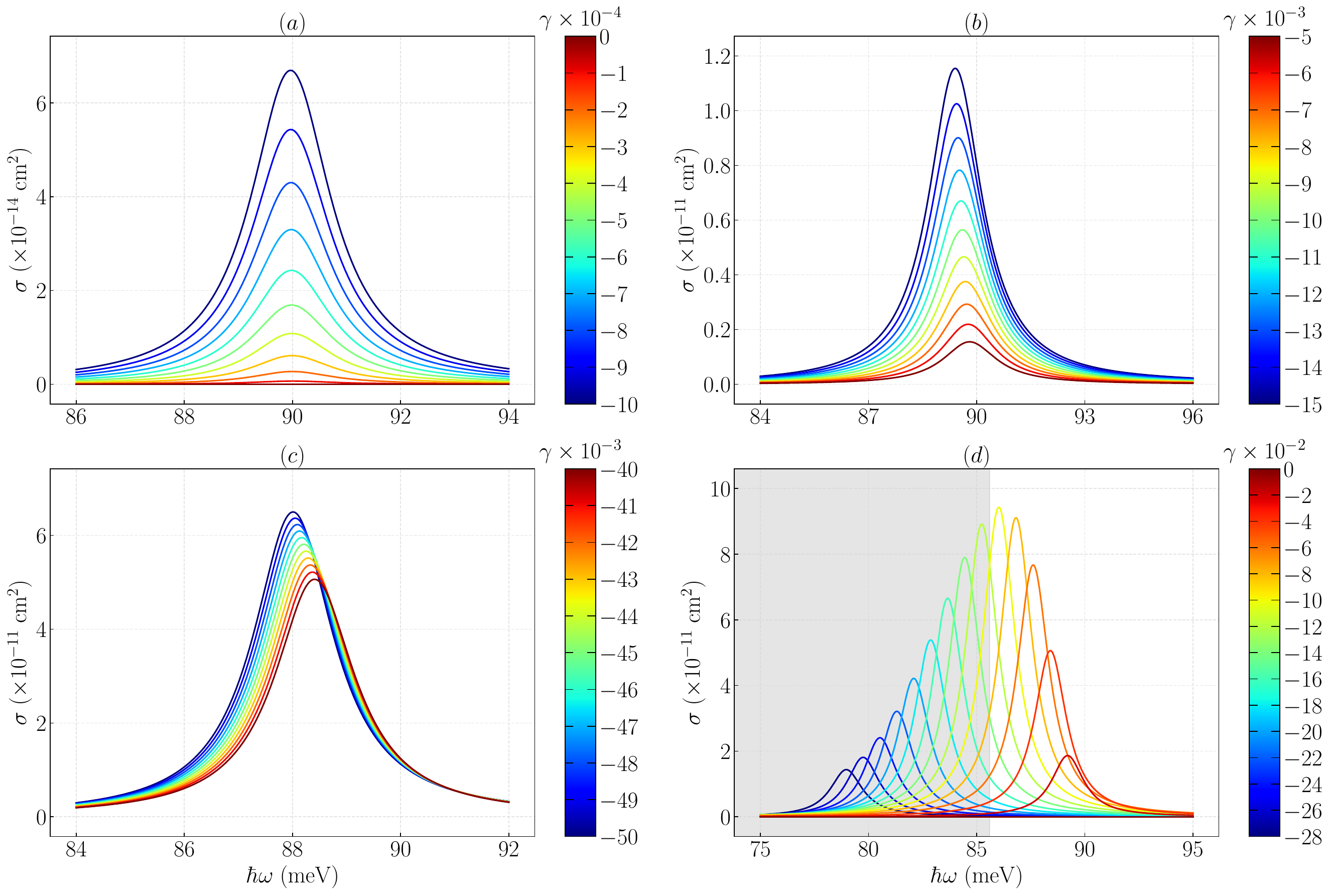}
\caption{Photoionization cross section only for negative values of $\gamma$ on transition $(n=0,\,\ell=0)$ to $(n=1,\,\ell=1)$. In (a), $\gamma$ varies from -10 to 0; in (b), from -15 to -5; in (c), from -50 to -40; and in (d), from -28 to 0. The scales are, respectively, $10^{-4}$, $10^{-3}$, $10^{-3}$, and $10^{-2}$. The physical parameters used are $\mu = 0.067 \,m_e$ and $\hbar \omega_0 = 30\,\text{meV}$.}
\label{fig:sigma3}
\end{figure}
\section{Photoionization Cross-Section}
\label{PCS}

The PCS, denoted as $\sigma$, quantifies the probability of an electron transitioning to a continuum state upon absorbing a photon. In essence, it describes the likelihood of ionizing electrons from a bound state under external optical excitation, heavily influenced by the confinement potential and the energy of the photons involved, which provides the mathematical expression defining the PCS \cite{lax1954,PRB.2008.77.045317}:
\begin{equation}
\sigma(\hbar \omega, \gamma, \omega_0) =\left(\frac{\zeta_{\mathrm{eff}}}{\zeta_0}\right)^2\frac{n_{r}}{\epsilon} \frac{4\pi^2}{3} \alpha_\text{fs} \hbar \omega |M_{fi}|^2 L(\hbar \omega)\;,
\end{equation}
where the ratio $\xi_{\text{eff}} / \xi_0$ represents the relationship between the effective electric field $\xi_{\text{eff}}$ of the incident photon and the average field $\xi_0$ in the medium. 
\begin{figure}[t]
    \centering
    \includegraphics[scale=0.45]{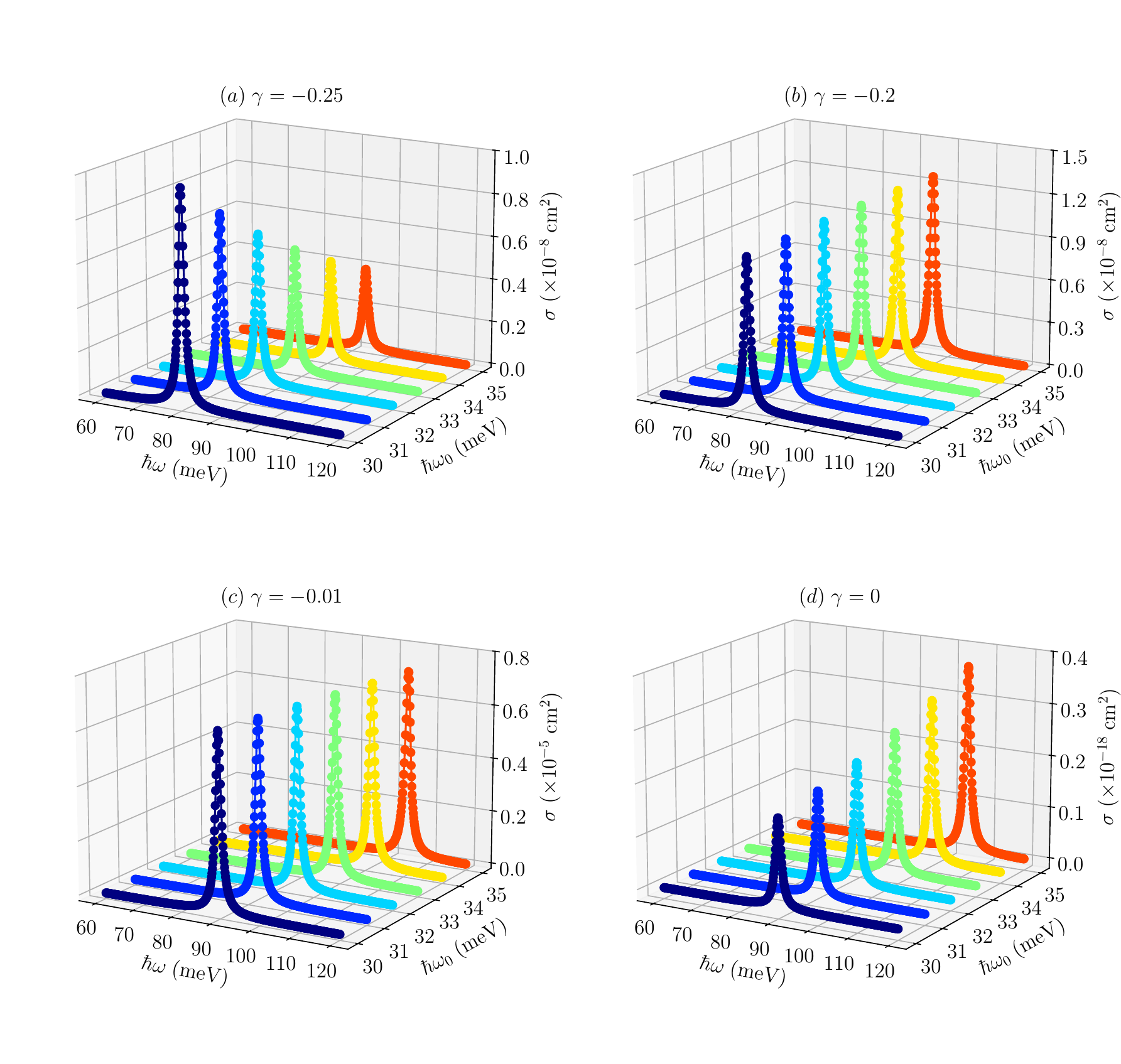}
    \caption{Photoionization cross section for different values of $\hbar \omega_0$ on transition $(n=0,\,\ell=0)$ to $(n=1,\,\ell=1)$. The physical parameters used are $\mu=$ $0.067\, m_e$ and (a) $\gamma=-0.25$, (b) $\gamma=-0.2$, (c) $\gamma=-0.01$ and (d) $\gamma=0$.}
    \label{fig:pcs3d0011}
\end{figure}
In this work, we will consider this ratio equal to 1 since it does not affect the shape of the cross-section \cite{SM.2013.58.94}, $\hbar \omega$ is the photon energy, $\Gamma_{fi}$ is the decay rate, $\alpha_\text{fs}$ is the fine-structure constant, $M_{fi}$ is the transition matrix element, $E_{fi}$ is the energy difference between the initial and final states, and $L(\hbar \omega)$ is the Lorentzian function, given by
\begin{equation}
    L(\hbar \omega) = \frac{\hbar\Gamma_{fi}}{\pi \left[(E_{fi} - \hbar \omega)^2 + (\Gamma_{fi} \hbar)^2\right]},
\end{equation}
which accounts for the natural broadening of spectral lines.

The transition matrix element is calculated as follows: 
\begin{equation} M_{if} = \int_0^\infty \psi_{n,\ell}^{(i)}(r) \psi_{n,\ell}^{(f)}(r) r^3 dr, \end{equation} 
where the functions $\psi_{n,\ell}^{(i)}(r)$ and $\psi_{n,\ell}^{(f)}(r)$ represent, respectively, the wave functions of the initial and final quantum states involved in the transition. The nonzero elements of the matrix $M_{if}$ follow the selection rule $\Delta\ell = \pm 1$ \cite{CTP.2024.76.105701,quantum6040041}. This integral incorporates both the overlap of the wave functions and the dependence of the effective mass on position.

Figure \ref{fig:matrixM} shows the plots of the matrix elements $\left|M_{if}\right|^2$ as a function of $\hbar\omega_{0}$ for two distinct transitions: $(\psi_{00} \rightarrow \psi_{11})$ in (a) and $(\psi_{01} \rightarrow \psi_{02})$ in (b), represented respectively by square and circular markers. In this study, only negative values of $\gamma$ are considered. The transition matrix directly influences the peak amplitude in the photoionization cross-section. It is observed that the values of $\left|M_{21}\right|^2$ are consistently higher than those of $\left|M_{10}\right|^2$. Moreover, as $\gamma$ varies, the elements $\left|M_{if}\right|^2$ change: as $\gamma$ increases, approaching zero, the values of $\left|M_{10}\right|^2$ decrease, while those of $\left|M_{21}\right|^2$ increase. This behavior highlights how small variations in the parameter $\gamma$ can significantly affect the transition matrix values.

\subsection{Numerical Results}

Now, we investigate the PCS numerically for various values of $\gamma$ and $\omega_0$. Namely, the photon energy $\hbar \omega$ is varied to capture the resonance peaks corresponding to specific transitions. Here, all calculations were performed using the following physical parameters and constants: $\mu=0.067 m_{e}$, where $m_{e}$ is the free electronic mass, $n_r=3.2$, $\epsilon= n_{r}^2 \, \epsilon_0$, $\epsilon_{0}=8.854\times 10^{-12}$ F/m, $\alpha_{fs}=1/137$ and $\hbar \Gamma_{fi}=0.9$ meV \cite{AdP.2012.524.327,OC.2011.284.5818}.

Figure \ref{fig:sigma1} illustrates the impact of the $\gamma$ factor on the photoionization cross-section profile. We consider the transition $\psi_{01} \to \psi_{02}$, covering both positive and negative $\gamma$ values on the order of $10^{-2}$, including the $\gamma = 0$. In graphs (a), (b), and (c), it is observed that for $\gamma = 0$ – when the electron mass remains constant throughout the region – the cross-section peak is centered at $\hbar\omega = 30~\mathrm{meV}$, which coincides with the value adopted for the analysis, $\hbar\omega_{0} = 30~\mathrm{meV}$.
\begin{table*}[!h]
\caption{Peak shifts and corresponding numerical values of PCS peaks for different values of $\gamma$ on transition $(n=0,\,\ell=1)$ to $(n=0,\,\ell=2)$.}
\begin{center}
\begin{tabular}{ccc}
\hline\hline
$\gamma \times 10^{-2}$ & $\hbar \omega\,(\mathrm{meV})$ & Peak of $\sigma\,\left(\times 10^{-5} \mathrm{~cm}^2\right)$ \\
\hline
-12 & 29.820 & 0.032 \\
-8 & 29.820 & 0.073 \\
-4 & 29.940 & 0.160 \\
0 & 30.060 & 0.340 \\
4 & 30.060 & 0.705 \\
8 & 30.180 & 1.421 \\
12 & 30.301 & 2.775 \\
16 & 30.301 & 5.330 \\
20 & 30.421 & 9.693 \\
24 & 30.421 & 11.083 \\
28 & 30.541 & 4.065 \\
32 & 30.541 & 0.534 \\
36 & 30.661 & 0.038 \\
40 & 30.661 & 0.002 \\
\hline\hline
\end{tabular}
\label{tab:table0102}
\end{center}
\end{table*}
Furthermore, it is noted that as $\gamma$ increases, the peaks shift toward higher $\hbar\omega$ values; conversely, for lower $\gamma$ values, the peaks shift toward lower $\hbar\omega$ values. These shifts are subtly small. Table \ref{tab:table0102} shows these shifts for some values of $\gamma$, along with the numerical values of the corresponding PCS peaks. It is noted that for this transition, the shifts of the peaks on the $\hbar\omega$-axis are very small, while the PCS amplitudes undergo a significant change.

Another noteworthy point is that increasing $\gamma$ also raises the amplitude of the cross-section in graphs (a), (b), and (c). However, graph (d) of Fig. \ref{fig:sigma1} reveals an interesting behavior: for specific ranges of $\gamma$, as it increases, the peaks shift to lower photon energies and the peak amplitude decreases, indicating a reversal of this behavior at a certain point. Moreover, the order of magnitude in graph (d), $\sigma\times10^{-18}$, is much lower than that of the others, which exhibit $\sigma\times10^{-6}$ in (a) and $\sigma\times10^{-5}$ in (b) and (c), demonstrating that the probability of the optical transition occurring for these values of $\gamma$ is very small.

Figure \ref{fig:pcs3d0102} presents the PCS as a function of the incident photon energy, $\hbar\omega$, for different values of ~$\hbar \omega_0\in 30,31, \ldots, 35\, \mathrm{meV}$. Considering the optical transition $\psi_{01} \rightarrow \psi_{02}$, the graphs display the results for $\gamma = -0.25$ in (a), $\gamma = -0.025$ in (b), $\gamma = 0$ in (c), and $\gamma = 0.5$ in (d). As $\hbar\omega_0$ increases, the PCS peak shifts to higher values of the incident photon energy, indicating that a higher photon energy is required for the optical transition to occur. Comparing Figs. \ref{fig:pcs3d0102}(a) and \ref{fig:pcs3d0102}(b), one notes that when $\gamma$ increases from $-0.25$ to $-0.025$, the peak values also rise, suggesting a higher probability of the optical transition occurring. In Fig. \ref{fig:pcs3d0102}(c), corresponding to the standard case in which the electron effective mass is position-independent, the peak amplitudes remain quite similar for the various values of $\hbar\omega_0$. It is also observed that the peaks occur at $\hbar\omega = \hbar\omega_0$ for each corresponding PCS. Finally, in Fig. \ref{fig:pcs3d0102}(d), where $\gamma = 0.5$, the differences between the peaks become more pronounced. As $\hbar\omega_0$ increases, the peaks shift to higher PCS values, demonstrating that the probability of the optical transition occurring is greater for these $\hbar\omega_0$ values.

To explore the scenario in which the PCS involves the ground state, we consider the optical transition between the quantum states $\psi_{00} \rightarrow \psi_{11}$. Unlike the previous transition, $\psi_{01} \rightarrow \psi_{02}$, this analysis imposes a constraint on the values of $\gamma$, as discussed in the section on the effective potential. Specifically, the quantum state $(n=0, \ell=0)$ does not generate a bound state for $\gamma > 0$, restricting our study to $\gamma \leq 0$ values. This implies that the analysis is limited to the scenario where the electron's mass decreases with increasing distance.

In Fig. \ref{fig:sigma3}, four graphs of the PCS as a function of the incident photon energy are presented, exploring the parameter $\gamma$ for the transition $\psi_{00} \rightarrow \psi_{11}$. In Fig. \ref{fig:sigma3}(a), the values of $\gamma$ are very close to zero, on the order of $10^{-4}$, indicating that the peaks are almost centered at $\hbar\omega = 90\,\mathrm{meV}$. Moreover, as the values of $\gamma$ decrease, the amplitude of the PCS increases, indicating a higher probability for the optical transition to occur.

In Fig. \ref{fig:sigma3}(b), for an interval of $\gamma$ farther from zero, on the order of $10^{-3}$, it is observed that the peaks shift to lower photon energies as $\gamma$ decreases. In Fig. \ref{fig:sigma3}(c), considering a similar order of magnitude ($-\gamma \sim 10^{-3}$), it is verified that the amplitudes of the peaks become more similar and undergo a shift to lower values of the incident photon energy.

An interesting case occurs in Fig. \ref{fig:sigma3}(d), where the values of $\gamma$ are considerably larger, on the order of $10^{-2}$. In this figure, a gray-shaded region highlights an interval in which the peaks increase in amplitude and shift to energies higher than $\hbar \omega$ as $\gamma$ increases, exhibiting behavior opposite to that observed in the other three graphs for this transition. However, the peaks outside this region decrease, indicating a lower probability of the optical transition. This phenomenon suggests the existence of an inversion. There is an interval in which the particle with higher inertia has a higher probability of undergoing an optical transition between states. In contrast, this probability is more remarkable for particles with lower inertia in another interval.

In Figure \ref{fig:pcs3d0011}, we analyze the transition $\psi_{00} \rightarrow \psi_{11}$ for different values of $\hbar\omega_{0}$, using the same parameters as in Figure \ref{fig:pcs3d0102}. For this transition, starting from the ground state, values of $\gamma$ were chosen to ensure the existence of a bound state.  In Figure \ref{fig:pcs3d0011}(a), with $\gamma = -0.25$, it is observed that the PCS peaks decrease rapidly as $\hbar\omega_{0}$ increases. This indicates that the transition probability from the $(n=0,\,\ell=0)$ state to the $(n=1,\,\ell=1)$ state, during the absorption of a photon by the quantum dot, is higher for lower values of $\hbar\omega_{0}$ within the analyzed range.  When slightly increasing the value of $\gamma$ to $\gamma = -0.2$, as shown in Figure \ref{fig:pcs3d0011}(b), the PCS peaks gradually increase as $\hbar\omega_{0}$ grows. This behavior demonstrates that a small variation in $\gamma$ causes a significant change in the photoionization cross-section profiles.  In Figure \ref{fig:pcs3d0011}(c), for $\gamma = -0.1$, the amplitudes of the PCS peaks show little variation among themselves. Finally, in Figure \ref{fig:pcs3d0011}(d), for $\gamma = 0$, a condition corresponding to the optical transition of a standard harmonic oscillator, where the particle mass is constant, the PCS peaks begin to increase with $\hbar\omega_{0}$ within the analyzed range. This behavior indicates that the probability of an optical transition increases as $\hbar\omega_{0}$ grows.

\section{Conclusions}
\label{conc}

In conclusion, this study provided insights into the optical properties of quantum dots with position-dependent effective mass. The analytical and numerical framework developed here enables computations. It enhances the understanding of the photoionization process, which is essential for gaining deeper insights into how QDs interact with electromagnetic radiation. Furthermore, these results can be generalized to other quantum systems exhibiting similar inhomogeneous material properties. Our study demonstrated that the particle's effective mass variation with position significantly alters the PCS profiles. Different choices of the parameter $\gamma$, which controls the spatial variation of mass, directly affect the amplitude and behavior of the PCS peaks. 

We analyzed two distinct scenarios for different values of $\gamma$: In the first case, the ground state participates in the optical transition, requiring $\gamma$ to take negative values. In the second case, the ground state does not participate in the transition, and $\gamma$ can take any value, provided that $\gamma \neq -2$. The results indicate that the intensity of PCS peaks can increase or decrease depending on the $\hbar\omega_{0}$ regime, suggesting that the localization of the bound state influences the transition probability. 

We believe that the results of our study can be applied to semiconductor devices and optical sensors based on quantum dots, leveraging the cross-section modification due to mass dependence. Controlling the effective mass can be a strategic mechanism to tune nanostructure optical transitions selectively. 

\section*{Acknowledgement}

This work was partially supported by the Brazilian agencies CAPES, CNPq,
FAPES, and FAPEMA. Edilberto O. Silva acknowledges the support from the
grants CNPq/306308/2022-3, FAPEMA/UNIVERSAL-06395/22, FAPEMA/APP-12256/22.
This study was partly financed by the Coordenação de
Aperfeiçoamento de Pessoal de Nível Superior - Brazil (CAPES) -
Code 001. F. S. Azevedo acknowledges CNPq Grant No. 153635/2024-0. C. Filgueiras acknowledge FAPEMIG Grant No. APQ 02226/22. and CNPq
Grant No. 310723/2021-3.

\section*{\label{sec:author}Author contributions}

All authors contributed equally to the paper.

\section*{\label{sec:datar}Data Availability Statement}

Data will be made available on reasonable request.

\bibliographystyle{model1a-num-names}
%\bibliography{mybibliography}

\end{document}